\documentclass[aps,pre,twocolumn,amsmath,amssymb,showpacs,amsfonts]{revtex4}
\usepackage{epsfig}
\usepackage{graphicx}
\usepackage{dcolumn}
\usepackage{bm}

\begin{document}

\title{Distribution of particles and bubbles in turbulence at small Stokes number}

\author{Itzhak Fouxon}
\affiliation{Raymond and Beverly Sackler School of Physics and Astronomy,
Tel-Aviv University, Tel-Aviv 69978, Israel}
\begin{abstract}

The inertia of particles driven by the turbulent flow of the surrounding fluid makes them prefer certain regions of the flow.
The heavy particles lag behind the flow and tend to accumulate in the regions with less vorticity, while the light particles do the opposite.
As a result of the long-time evolution, the particles distribute over a multi-fractal attractor in space. We consider this distribution using our
recent results on the steady states of chaotic dynamics. We describe the preferential concentration analytically and
derive the correlation functions of density and the fractal dimensions of the attractor. The results are obtained for real turbulence and are testable experimentally.

\end{abstract}
 \maketitle

Recently the problem of distribution of inertial particles in turbulence received much attention of the researchers \cite{MaxeyRiley,Maxey,BFF,FFS1,FalkovichPumir,BecGawedzkiHorvai,Collins,BecCenciniHillerbranddelta,Stefano,Cencini,Olla,MehligWilkinson}.
This is largely thanks
to the breakthrough in the theoretical understanding of the Lagrangian motion of particles in the flow that occurred lately \cite{review}.
While the understanding of the behavior of particles that have negligible inertia and follow the flow is quite complete by now
\cite{review}, the understanding of the behavior of inertial particles is still insufficient. This is while the subject has an
extremely wide range of applications: the flows of fluids are typically turbulent and often laden with external particles. Theoretical advancement was made mainly for the case of small Stokes
number, where the inertia is weak and the particles "almost" follow the flow. Even in this limit of small Stokes numbers, the particles' distribution is highly
non-trivial. Particles' deviations from the surrounding flow accumulate with time, bringing particles to a strange attractor in
space. This attractor is multi-fractal and the only theoretical result obtained so far for the real turbulent flow was the derivation of the correlation codimension \cite{FFS1}. Here a result obtained for real turbulence is a result obtained without modeling turbulence and expressed in terms of the (unknown) statistical properties of turbulence. Since the statistics of
turbulence is largely unknown \cite{Frisch}, then to obtain such a result one needs to make universal predictions on particles' behavior in the flow
independent of the details of the statistics of that flow.

In this Letter we provide the complete description of the distribution of particles in real turbulence at small Stokes numbers, describing both the correlation of the particles' density with the surrounding flow and the statistics of the singular density on the
attractor. We give a number of predictions that are testable experimentally.

The idea that particles' inertia leads to inhomogeneous spatial distribution dates back to the seminal paper by Maxey \cite{Maxey}. It was observed
that due to inertia heavy particles are pushed out of the vortices and hence they will not distribute uniformly in the flow, like the
inertia-less particles. However the quantitative description of the correlations between the locations of particles and of vortices stayed unaddressed. Note the distribution of vorticity in turbulence is random and dynamical, while the distribution of particles reflects its cumulative effect over time. There is a residual correlation that we describe by an integral relation holding in the steady state.

We find the spectrum of fractal dimensions of the attractor. We show that while the correlation dimension is different
from the dimension of space, the fractal or similarity dimension \cite{HentshelProccacia} is equal to the space dimension. In contrast the
information dimension is different from the spatial dimension and it equals the Kaplan-Yorke dimension. In turn, the correlation codimension equals twice the Kaplan-Yorke codimension which constitutes a prediction allowing direct testing in the laboratory.

The analysis is based on the recent finding of a universal description for the steady state density of the weakly compressible dynamical systems \cite{Tzhak}. The particles' motion, though governed by Newton's law, admits an effective description in terms of a velocity field in space. Inertia is described by a small compressible correction to the incompressible velocity of the background turbulent flow. This correction leads to a small disbalance of trajectories going in and out of space regions, which accumulates over a long time to a big effect. Thus compressibility is a singular perturbation which treatment was performed in \cite{Tzhak}.
For a mixing incompressible velocity the evolution of a small volume of particles makes it dense in space. The volume's coarse-graining over an
arbitrarily small scale covers all the available space, which volume is assumed finite. When a small compressible component is added to the velocity, the coarse-graining of the evolved volume over an arbitrarily small scale does not cover the whole space any longer.
However the coarse-graining over a small but finite scale, that tends to zero with compressibility, already covers the whole volume.

The analysis assumes the single-particle approximation where one neglects the interaction between the particles and their back reaction on the flow. We consider a small spherical particle with the radius $a$ and the material density $\rho_p$ suspended in a fluid with the density $\rho$ and the kinematic viscosity $\nu$. The fluid flow
$\bm u(t, \bm x)$ is assumed to be incompressible. The Newton law governing the evolution of the particle's position $\bm x(t)$ and the particle's velocity $\bm v(t)$ is assumed to have the form
\begin{eqnarray}&&
\frac{d\bm v}{dt}=\gamma \frac{d}{dt}\bm u[t, \bm x(t)] -\frac{\bm v-\bm u[t, \bm x(t)]}{\tau},
\end{eqnarray}
where $\gamma = 3\rho/(\rho + 2\rho_p)$ and $\tau = a^2/(3\nu \gamma)$ is the Stokes time. Thus we assume that all the forces besides
the added mass and the drag can be neglected \cite{Cencini,Maxey}. With no loss we set the total volume and the mass equal to one, so the spatial average of the particles' density $n$ obeys $\langle n\rangle=1$. We set $\beta=\gamma-1$ so the particle's velocity relative to the flow $\bm w(t)\equiv \bm v(t)-\bm u[t, \bm x(t)]$ obeys
\begin{eqnarray}&&
\frac{d\bm w}{dt}=-\frac{\bm w}{\tau}+\beta \frac{d}{dt}\bm u[t, \bm x(t)].
\end{eqnarray}
The parameter $\beta=2(\rho-\rho_p)/(\rho + 2\rho_p)$ changes from $-1$ for heavy particles to $2$ for light ones. After transients
\begin{eqnarray}&&
\bm w(t)=\beta \int_{-\infty}^t \exp\left[-\frac{t-t'}{\tau}\right]\frac{d}{dt'}\bm u[t', \bm x(t')]dt'.
\end{eqnarray}
We assume $\tau$ is much smaller than the smallest time-scale of turbulence, which is the viscous time-scale $t_{\eta}$, so
the Stokes number $St\equiv \tau/t_{\eta}\ll 1$. Then we can substitute the derivative in the integrand by its value at $t'=t$ so
$\bm v(t)\approx \bm u+\mu\left[\partial_t\bm u+(\bm u\cdot \nabla)\bm u\right]$ with $\mu\equiv \beta\tau=2a^2(\rho\!-\!\rho_p)/(9\nu\rho)$.
%\begin{eqnarray}&&
%\!\!\!\!\!\!\!\!\!\!\!
%\bm v(t)\!=\!\bm u\!+\!\mu\left[\partial_t\bm u\!+\!(\bm u\cdot \nabla)\bm u\right],\ \ \mu\equiv \beta\tau\!=\!\frac{2a^2(\rho\!-\!\rho_p)}{9\nu\rho}.
%\end{eqnarray}
%where the RHS is evaluated at the particle's position $\bm x(t)$.
Thus at $St\ll 1$ the particle's velocity is determined
uniquely by its position $\bm x(t)$ in space
and one can introduce the particle's velocity field $\bm v(t, \bm x)$
\begin{eqnarray}&&
\!\!\!\!\!\!\dot {\bm x}(t)=\bm v\left[t, \bm x(t)\right],\ \ \bm v\equiv \bm u+\mu\left[\partial_t\bm u+(\bm u\cdot \nabla)\bm u\right].\label{basic}
\end{eqnarray}
In the zero inertia limit ${\rm St}\to 0$ the particles follow the incompressible mixing flow of turbulence ${\dot {\bm x}}=\bm u\left[t, \bm x(t)\right]$ and in the steady state they are uniformly distributed in space, so their steady state density $n_s$ equals one. This behavior is characteristic of small dye particles. However, at a small but finite $St$, the small correction $\bm v-\bm u$ gives the particles' velocity field a finite compressibility \cite{Maxey}
\begin{eqnarray}&&\!\!\!\!\!\!\!\!\!\!\!\!\!\!
w\equiv \nabla\cdot \bm v=%(\beta-1)\tau \nabla\cdot [(\bm u\cdot\nabla)\bm u]=
-\mu\phi
%\tau\nabla^2 p,
\neq 0,\ \ \phi=\omega^2-s^2,\label{divergence}
\end{eqnarray}
so the constant is no longer a solution to the continuity equation $\partial_t n+\nabla\cdot(n\bm v)=0$. Above $s^2=s_{ij}s_{ij}$ and $\omega^2=a_{ij}a_{ij}$, where $s_{ij}$ is the symmetric (strain) and $a_{ij}$ is the antisymmetric (vorticity) parts of velocity gradient $\partial_j u_i=s_{ij}+a_{ij}$. The field $\phi(\bm x)$ is positive in the regions dominated by vorticity and negative in the regions dominated by the strain, and it will be called below the indicator, indicating whether $\bm x$ is in a vortex. It follows
from the Navier-Stokes equations that $\phi$ equals the Laplacian of the turbulent pressure, $\phi=\nabla^2 p$.
Eq.~(\ref{divergence}) shows that heavy particles $\beta<0$ are repelled from vortices (here and below "vortex" is used qualitatively), while the light ones $\beta>0$ are attracted. This is the generalization
of the familiar fact that a heavy particle in a centrifuge is pushed out to the boundary. Turbulence can be considered as a dynamically
changing spatial distribution of vorticity, so heavy particles tend to accumulate on the boundaries between the vortices, cf. \cite{MaxeyRiley,FFS1,Olla}, forming a singular density supported on these boundaries. This accumulation however is insignificant during the life-time %$\tau_l$
of a single vortex and the ultimate singular distribution of particles in space $n_s$ forms from the long-time %(much longer than $\tau_l$)
combined action of many uncorrelated vortices.
%This hints at the possibility of a universal description in the spirit of the central limit theorem, which is performed below.
Still one expects a residual correlation between the distributions of vorticity and particles, to find which we consider the steady state density $n_s$. One expects $n_s$ can be obtained by letting an arbitrary initial condition $n_0$ in the remote past $n(t=-T)=n_0$ evolve for
infinite time, $T\to\infty$, according to the continuity equation. Starting from the uniform initial distribution %$n_0=1$
we obtain the steady state density
\begin{eqnarray}&&\!\!\!\!\!\!\!\!\!
n_s(\bm x)\!=\!\!\lim_{T\to\infty}n(T),\ \ n(T)\!=\!\exp\left[\!-\!\int_{-T}^0\! w[t, \bm q(t, \bm x)]dt\right], \label{steady}
\end{eqnarray}
provided a condition of decay of correlations \cite{Tzhak}, that should hold in our case, is met. For the cross-correlation of density and vorticity $F(\bm x)=\langle \phi(0)n_{s}(\bm x)\rangle$ we find
\begin{eqnarray}&&
\!\!\!\!\!
F(\bm x)\!\!=\!\!\left\langle \phi(0, 0)\exp\left[\mu\!\int_{-\infty}^0\!\!\phi[t, \bm q(t, \bm x)]dt\right]\right\rangle
\label{exponent}
\\&&\!\!\!\!\!
=\partial_{\alpha}\ln \left\langle \exp\left[\alpha \phi(0, 0)+\mu\int_{-\infty}^0 \phi[t, \bm q(t, \bm x)]dt\right] \right\rangle|_{\alpha=0},\nonumber
\end{eqnarray}
where we used the conservation of the mean density $\langle n(t)\rangle=const$. Applying the cumulant expansion \cite{Ma}, taking derivative of the series and setting $\alpha=0$ we find %to the leading order in $St$
\begin{eqnarray}&&
\!\!\!\!\!F(\bm x)=\mu \int_{-\infty}^0 dt  \langle\langle \phi(0, 0)\phi[t, \bm q(t, \bm x)]\rangle\rangle_c+O(St^2). \label{asymptote}
\end{eqnarray}
The above formula is the same as one would obtain by expanding the exponent in Eq.~(\ref{exponent}) and keeping the lowest order term in $\tau$,
with one important difference. In Eq.~(\ref{asymptote}) one has the second order cumulant or dispersion, that one would not get by the series expansion of the
exponent, cf. \cite{Maxey}. This difference is essential as without the cumulant the integral in Eq.~(\ref{asymptote}) diverges: $\langle w[t, \bm q(t, \bm x)]\rangle=(1-\beta)\tau \langle \phi[t, \bm q(t, \bm x)]\rangle$ is equal to a non-zero sum of Lyapunov exponents, see below.
To leading order in $St$ one can substitute $\bm q(t, \bm x)$ in Eq.~(\ref{asymptote})
by $\bm X(t, \bm x)$
\begin{eqnarray}&&
\partial_t \bm X(t, \bm x)=\bm u[t, \bm X(t, \bm x)],\ \ \bm X(0, \bm x)=\bm x.
\end{eqnarray}
where $\bm X(t, \bm x)$ are Lagrangian trajectories of $\bm u$.
One finds
\begin{eqnarray}&&
\!\!\!\!\!\!\langle \phi(0)n_{s}(\bm x)\rangle=\mu\int_{-\infty}^0 dt  \langle \phi(0, 0)\phi[t, \bm X(t, \bm x)]\rangle,
\end{eqnarray}
where we can already omit the cumulant since by incompressibility $\langle w[t, \bm X(t, \bm x)]\rangle=\langle w(t, \bm x)\rangle$, while $\int w(t, \bm x)d\bm x=\int \nabla\cdot\bm v(t,\bm x)=0$ by the boundary conditions. Since there is no degeneracy, the non-negative spectrum of the Laplacian of pressure in the Lagrangian frame $\phi[t, \bm X(t, \bm x)]$ is strictly positive at zero frequency
\begin{eqnarray}&&
E(0)=\int_{-\infty}^{\infty} \langle \phi(0, 0)\phi[t, \bm X(t, 0)]\rangle>0.
\end{eqnarray}
The single-point correlation $\langle \phi n_{s}\rangle\equiv \int \phi(\bm x)n_s(\bm x)d\bm x$ equals $\mu E(0)/2$ and it gives the integral of $\phi$ where each region weighted by the number of particles in it
\begin{eqnarray}&&
\!\!\!\!\!\!\!\!\!\!\!\!\!\!\!\int \!\left[\omega^2(\bm x)\!-\!\!s^2(\bm x)\right]n_s(\bm x)d\bm x\!=\!a^2(\rho\!-\!\rho_p)(9\nu\rho)^{-1}E(0),\label{central}
\end{eqnarray}
where we used the definitions of $\phi$ and $\mu$. For heavy particles, $\rho_p>\rho$, the answer is negative giving a measure of the extent to which the particles favor regions with negative $\phi$. For light particles, $\rho_p<\rho$ the answer is positive measuring their favoring of vortices. The above integral steady state relation holds at any $t$.

The quantity $E(0)$ appeared first in \cite{FFS1}, where it was shown to determine the correlation dimension of the particles' attractor in space.
This quantity is increased by the intermittency of turbulence and it can be estimated as $t_{\eta}^{-3}f(Re)$ where $f(Re)$ is a growing function of the Reynolds number $Re$ that grows as a power \cite{FFS1,FalkovichPumir}. We show $E(0)$ determines all the fractal dimensions.

We observe that Eq.~(\ref{basic}) is a weakly dissipative dynamical system, defined as the dynamics for which the potential part of $\bm v$ is much smaller than the solenoidal one. The statistics of the steady state density of such systems was shown recently to allow for a complete and universal
description \cite{Tzhak}. The application to our case gives the following results.
The motion of particles in space is chaotic and is characterized by the Lyapunov exponents \cite{Oseledets}. To the lowest
order in $St$ the exponents are equal to the Lyapunov exponents $\lambda_i$ of the turbulent flow $\bm u$. However, the value of the sum of the Lyapunov exponents $\sum \lambda_i^+$ that determines the logarithmic rate of growth of the volumes forward in time,
\begin{eqnarray}&&
\sum \lambda_i^+(\bm x)\equiv\lim_{t\to\infty} t^{-1}\ln \det\nabla_j q_i(t, \bm x),
\end{eqnarray}
is zero for $\bm u$, so the leading order approximation demands the account of the correction $\bm v-\bm u$. This is also the case
of the sum of the Lyapunov exponents $\sum \lambda_i^-$ of the backward-in-time flow that determines the density \cite{Tzhak}
%(to find the density at $\bm x$ at $t=0$ one needs to track the trajectories from $\bm x$ back, rather than forward, in time)
\begin{eqnarray}&&
\lim_{T\to\infty}T^{-1}\ln n(0, \bm x, T)=\sum \lambda_i^-(\bm x).
\end{eqnarray}
For turbulence $\sum\lambda_i^{\pm}$ is expected to be the same for all $\bm x$ with the possible exception of a set of points with zero volume. The results of \cite{Tzhak,FF} give
\begin{eqnarray}&&
\!\!\!\!\!\!\!\sum \lambda_i^{\pm}\approx -\frac{1}{2}\int_{-\infty}^{\infty}\langle w(0, 0) w[t, \bm X(t, 0)] \rangle =-\frac{\mu^2 E(0)}{2}.\nonumber
\end{eqnarray}
Thus for all initial points, with the possible exception of a set of points with zero volume, the infinitesimal volumes decay to zero in the
limit of infinite evolution time, while the steady state density $n_s$ is zero except for a set of points with zero volume. Due to conservation of mass $\int n d\bm x$, we conclude that $n_s$ has $\delta-$function type singularities on its support. This support is the strange attractor - the multifractal set in space that is approached by the particles' trajectories at large times.
We now find the Kaplan-Yorke codimension $C_{KY}$ of the attractor. At weak compressibility the definition \cite{KY} of $C_{KY}$
reduces \cite{Tzhak} to $C_{KY}=\sum \lambda_i^+/\lambda_3^+$ which gives to the leading order
\begin{eqnarray}&&
C_{KY}=\frac{\mu^2 E(0)}{2|\lambda_3|}=\frac{2a^4(\rho\!-\!\rho_p)^2E(0)}{81\nu^2\rho^2|\lambda_3|}, \label{KY}
\end{eqnarray}
where the third Lyapunov exponent $\lambda_3$ determines the rate of exponential separation of $\bm X(t, \bm x)$ back
in time. We have $|\lambda_3|\sim t_{\eta}^{-1}$ and $C_{KY}\sim \beta^2 St^2 f(Re)$, cf. \cite{FFS1,Collins}.

The probability for two particles to be at the distance $\bm x$ is described by the pair-correlation function $\langle n_s(0)n_s(\bm x)\rangle$. Substituting for $n_s$ the expression from Eq.~(\ref{steady}) and using the cumulant expansion \cite{Tzhak} one finds $\langle n_s(0)n_s(\bm x)\rangle=\exp\left[\mu^2 g(\bm x)\right]$ where the structure function $g(\bm x)$ depends only on the statistics of turbulence
\begin{eqnarray}&&
g(\bm x)\equiv \int_{-\infty}^0\!\!\!\! dt_1dt_2\langle
\phi[t_1, {\bf X}(t_1, 0)] \phi[t_2, {\bf X}(t_2, {\bf x})]\rangle. \label{answer}\end{eqnarray}
The above is valid if the higher order terms in the cumulant expansion are negligible \cite{Tzhak}. The Kolmogorov theory (KT) estimate would give the validity condition $St\ll 1$, while the account of intermittency changes the condition to $h(Re)St\ll 1$ where $h(Re)$ is expected to be a slowly growing function of $Re$, cf. \cite{FalkovichPumir}. The function $g(\bm x)$ has a universal behavior \cite{Tzhak} at small $\bm x$ that gives
\begin{eqnarray}&&
\!\!\!\!\!\!\!\!\!\!\!\!\langle n_s(0)n_s(\bm x)\rangle=\left(\eta/x\right)^{2C_{KY}},\ \ x\ll \eta, \label{smallscales}
\end{eqnarray}
where $\eta\sim (\nu t_{\eta})^{1/2}$ is the Kolmogorov scale of turbulence \cite{FFS1}. Thus the correlation codimension equals $2C_{KY}$.

Remarkably, the structure  function determines all the correlation functions of $n_s$. Generalization of the calculation of $\langle n_s(0)n_s(\bm x)\rangle$ gives the log-normal statistics \cite{Tzhak}
\begin{eqnarray}&&
\!\!\!\!\!\!\!\!\!\!\!\!\!\!\!\!\!\!\!\!\langle n_{s}(\bm x_1)n_{s}(\bm x_2)..n_{s}(\bm x_N)\rangle%\left\langle \prod_{i=1}^N n_{SRB}({\bf r}_i)\right\rangle=\exp\left[
\!=\!\exp\left[\!\mu^2\!\sum_{i>j}g({\bf x}_{i}\!-\!{\bf
x}_j)\!\right], \label{bueno}\end{eqnarray}

The density $n_s$ does not have physical meaning and we consider the coarse-grained density $n_l$,
\begin{eqnarray}&&
m_l(\bm x)\equiv \int_{|\bm x'-\bm x|<l}n_s(\bm x')d\bm x',\ \ n_l(\bm x)\equiv 3m_l(\bm x)/(4\pi l^3), \nonumber
\end{eqnarray}
where $m_l(\bm x)$ is the mass in a small ball.
The limits $l\to 0$ and $St\to 0$ do not commute ($\langle n_l^2\rangle\sim (\eta/l)^{2C_{KY}}$)
\begin{eqnarray}&&
\lim_{l\to 0}\lim_{St\to 0} \langle n_l^2 \rangle=1,\ \ \lim_{St\to 0}\lim_{l\to 0}\langle n_l^2 \rangle=\infty,
\end{eqnarray}
For any $St>0$ the fluctuations of $n_l$ are large for a sufficiently small $l$. On the other hand, by $\lim_{St\to 0} \langle n_l^2 \rangle=1$
one sees that for any fixed $l>0$ the fluctuations of $n_l$ are small
for a sufficiently small $St$. The coarse-grained density is uniform over scales which minimal value vanishes with $St$.
Thus turbulence effect on the particles depends on the observer's resolution $l$: at $2C_{KY}\ln \left(\eta/l\right)\gtrsim 1$
segregation holds, while $2C_{KY}\ln \left(\eta/l\right)\ll 1$ - mixing. This is how mixing works effectively for particles on a multifractal. Segregation may also bring physical effects \cite{FFS1}.

At $St\ll 1$ there is a scale $L\ll \eta$ over which the density is almost uniform. We note that $m_l(t=0, \bm x)$ is equal to the mass contained in the preimage of the ball time $t$ ago, which is an ellipsoid around $\bm q(-t, \bm x)$ with the largest axis growing as
$l\exp[|\lambda_3 t|]$. At $t_*=|\lambda_3|^{-1}\ln(L/l)$ the ellipsoid has the scale over which the density is uniform, so the mass
contained in it is just its volume $4\pi l^3 \exp[-\int_{-t_*}^0 w[t', \bm q(t', \bm x)]dt']/3$ and we find
\begin{eqnarray}&&
\!\!\!\!\!\!\!\!\!\!\!\!\!\!n_l(\bm x)=\exp\left[-\mu \int_{-|\lambda_3|^{-1}\ln(L/l)}^0 \phi[t', \bm q(t', \bm x)]dt'\right],
\end{eqnarray}
see \cite{Tzhak} for details.
%The above formula gives a fundamental physical insight into the fluctuations of particles' density at $St\ll 1$. The fluctuations of $n_l$ at small scales are created by the condensation of mass from a volume of size much smaller than $\eta$ over which the density is roughly uniform.
The smallness of $\mu$ brings the expected conclusion that the statistics of $n_l$ is log-normal
\begin{eqnarray}&&
\langle n_l^{\rho}\rangle=\left(\eta/l\right)^{C_{KY}\rho(\rho-1)},
\end{eqnarray}
cf. Eq.~(\ref{bueno}). %For $\gamma=2$ the above formula reproduces the previous answer for the pair correlation function.
The spectrum of the fractal dimensions $D(\alpha)\equiv\lim_{l\to 0}\ln \langle m_l^{\alpha-1}n_{s}\rangle/[(\alpha-1)\ln l]$ %\label{dimensions}
%\end{eqnarray}
involves the average with $n_s$, rather than the spatial average \cite{HentshelProccacia,BecGawedzkiHorvai}. To find it
consider $\langle n_l^{\alpha-1}n_{s}\rangle=\lim_{T\to\infty}\!
\langle \exp[-\alpha\int_{-t_*}^0\!\!\omega[t, \bm q(t, \bm r)]dt\!-\!\int_{-T}^{-t_*}\!\!\omega[t, \bm q(t, \bm r)]dt]\rangle$.
Due to $St\ll 1$ the contribution of time-intervals with length $t_{\eta}$ is negligible and we may substitute the upper limit
in the last integral by $-t_*-t_{\eta}$ which allows to perform independent averaging
$\langle \exp[-\alpha\int_{-t_*}^0\!\!\omega[t, \bm q(t, \bm r)]dt\!-\!\int_{-T}^{-t_*-t_{\eta}}\!\!\omega[t, \bm q(t, \bm r)]dt]\rangle
\approx \langle \exp[-\alpha\int_{-t_*}^0\!\!\omega[t, \bm q(t, \bm r)]dt]\rangle \langle \exp[-\!\int_{-T}^{-t_*-t_{\eta}}\!\!\omega[t, \bm q(t, \bm r)]dt]\rangle$. However the last average is equal to one by the conservation of mean density, so $\langle n_l^{\alpha-1}n_{s}\rangle=\langle n_l^{\alpha}\rangle$ and
\begin{eqnarray}&&
D(\alpha)=3-C_{KY}\alpha. \label{dimensions222}
\end{eqnarray}
Our results generalize to the two-dimensional case, where they can be compared with \cite{BecGawedzkiHorvai}. Working out the small compressibility limit reproduces our answer.
%who relate $D(\alpha)$ to the statistics of the finite-time Lyapunov exponents.
Returning to the three-dimensional case, we observe that the fractal dimensions are close to $3$ (we do not consider $\alpha\gg 1$)
 %and
%the high moments of $n_l$ where the lognormal approximation is not valid and the compressibility of $\bm v$ cannot be considered small, cf. \cite{BecGawedzkiHorvai}.
The fractal dimension of the attractor $D(0)$ coincides with the space dimension $3$, which is somewhat counter-intuitive since the volume of the attractor is zero. The information dimension $D(1)$ is equal to the Kaplan-Yorke dimension.

At ${\rm St}\ll 1$ density inhomogeneities are absent in the inertial range \cite{Frisch}. Then Eq.~(\ref{smallscales}) is a complete description. In contrast, at ${\rm St}\sim 1$ the inertial range inhomogeneities are important \cite{Stefano} and Eq.~(\ref{answer}), extended to hold asymptotically at ${\rm St}\sim 1$, gives a unique access to the inhomogeneities.
In KT $g(\bm x)$ depends only on $x$ and the mean energy dissipation $\epsilon$
so $\langle n_s(0)n_s(\bm x)\rangle=\exp\left[C\mu^2\epsilon^{2/3}x^{-4/3}\right]$. This prediction describes correctly the model of ${\bf v}$ decorrelated in time \cite{BFF,Collins,BecCenciniHillerbranddelta}. However, for turbulence, simulations \cite{Stefano}
show $\ln \langle n_s(0)n_s(\bm x)\rangle\propto x^{-10/3}$ at moderate ${\rm Re}$ where KT is expected to work.
%The Reynolds numbers in \cite{Stefano} are moderate so significant difference from KT calls for a study.
Noting $g(\bm x)\sim \tau_x^2 \partial_x^4\langle
[p({\bf x})-p(0)]^2\rangle$, where $\tau_x$ is the relevant time-scale, we suggest the difference has the same origin as the deviations of the pressure scaling from KT \cite{Gotoh}.

%Throughout this work we considered relatively small density where the particles' back reaction on the flow and their hydrodynamic interactions are
%negligible.
%The results can be tested either considering the density of a dilute gas of particles, or the probability density function (PDF) of the coordinate of a single particle.

A central result is the analytic description of the preferential concentration by Eq.~(\ref{central}).
%the integral relation
%\begin{eqnarray}&&
%\int \left[\omega^2(\bm x)-s^2(\bm x)\right]n_s(\bm x)d\bm x=a^2(\rho-\rho_p)E(0)/(9\nu\rho).\nonumber
%\end{eqnarray}
A single number $E(0)$ completely characterizes the influence of turbulence on the log-normal statistics of density at $r\ll \eta$.
Log-normality arises because the steady state density
is the cumulative result of the creation of inhomogeneities by many uncorrelated vortices, each of which creates but weak inhomogeneity.
The fractal structure at scale $l$ forms relatively fast - within the characteristic time-scale $|\lambda_3|^{-1}\ln(\eta/l)$. The predictions are testable.

The author is grateful to J. Bec, M. Cencini, G. Falkovich, K. Gawedzki, J. Kurchan, A. Leshansky, R. Vilela, and M. Wilkinson, for discussions.
This work was supported by COST Action MP$0806$.


\begin{references}

\bibitem{MaxeyRiley} M. R. Maxey and J. J. Riley, Phys. Fluids {\bf 26}, 883 (1983).

\bibitem{Maxey} M. R. Maxey, J. Fluid Mech. {\bf 174}, 441 (1987).

\bibitem{BFF} E. Balkovsky, G. Falkovich and A. Fouxon, %ArXiv:chao-dyn/9912027;
Phys. Rev. Lett. {\bf 86}, 2790 (2001).

\bibitem{FFS1} G. Falkovich, A. Fouxon and M. Stepanov, Nature {\bf 419}, 151 (2002).

\bibitem{FalkovichPumir} G. Falkovich and A. Pumir, Phys. Fluids {\bf 16}, L47 (2004).

\bibitem{BecGawedzkiHorvai} J. Bec, K. Gawedzki, and P. Horvai, Phys. Rev. Lett. \textbf{92},
224501 (2004).

\bibitem{Collins} J. Chun, D. L. Koch, S. L. Rani, A. Ahluwalia, and L. R. Collins, J. Fluid Mech. {\bf 536}, 219 (2005).

\bibitem{BecCenciniHillerbranddelta} J. Bec, M. Cencini, and R.
Hillerbrand Phys. Rev. E {\bf 75}, 025301 (2007).

\bibitem{Stefano} J. Bec, L. Biferale, M. Cencini, A. Lanotte, S.
Musacchio, and F. Toschi, Phys. Rev. Lett. {\bf 98}, 084502 (2007).

\bibitem{Cencini} E. Calzavarini, M. Cencini, D. Lohse, and F, Toschi, Phys. Rev. Lett. {\bf 101}, 084504 (2008).

\bibitem{Olla} P. Olla, Phys. Rev. E \textbf{81}, 016305 (2010).

\bibitem{MehligWilkinson} M. Wilkinson, B. Mehlig and K. Gustavsson, Europhys. Lett. {\bf 89}, 50002 (2010).

%\bibitem{EqMotion} J. J. Bluemink, D. Lohse, A. Prosperetti, and L. Van Wijngaarden, J. Fluid Mech. {\bf 643}, 1-31 (2010).

\bibitem{review} G. Falkovich, K. Gawedzki, and M. Vergassola, Rev. Mod. Phys. {\bf 73}, 913–975 (2001).

\bibitem{Frisch} U. Frisch, {\it Turbulence: The Legacy of A. N.
Kolmogorov}, Cambridge Univ. Press, 1995.

\bibitem{HentshelProccacia} H. G. E. Hentschel and I. Procaccia, Phys. D \textbf{8}, 435 (1983).

\bibitem{Tzhak} I. Fouxon, 	arXiv:1110.1625.

\bibitem{Ma} S.-K. Ma, {\it Statistical Mechanics}, World Scientific Publishing Company (1984).

\bibitem{FF}  G. Falkovich and A. Fouxon, N. J. Phys. {\bf 6}, 50 (2004).% and complete version in arXiv:nlin/0312033.

\bibitem{Oseledets} V. I. Oseledets, Trans. Mosc. Math. Soc. {\bf 19}, 197 (1968).

\bibitem{KY} J. L. Kaplan and J. A.  Yorke, {\it Functional Differential Equations and Approximations of Fixed Points}, Ed. H. O. Peitgen and H. O. Walther, Springer, 204 (1979).

\bibitem{Gotoh} T. Gotoh and D. Fukayama, Phys. Rev. Lett. \textbf{86}, 3775 (2001).

\end{references}
\end{document}